\documentclass{moriond}

\usepackage{float}

\def\be{\begin{equation}}
\def\ee{\end{equation}}
\def\bea{\begin{eqnarray}}
\def\eea{\end{eqnarray}}

\usepackage{amsmath}
\usepackage{amsfonts}
\usepackage{amssymb}
\usepackage{mathrsfs}
\usepackage{graphicx}
\usepackage[dvipsnames]{xcolor}
\usepackage{color}
\usepackage[normalem]{ulem}
\usepackage{soul}
\setstcolor{red}
\usepackage{subfig}

\newcommand{\eg}{{\it e.g.}}

\newcommand{\eq}{Eq.}
\newcommand{\eqs}{Eqs.}

\newcommand{\fig}{Fig.}

\newcommand{\Photo}{\includegraphics[height=35mm]{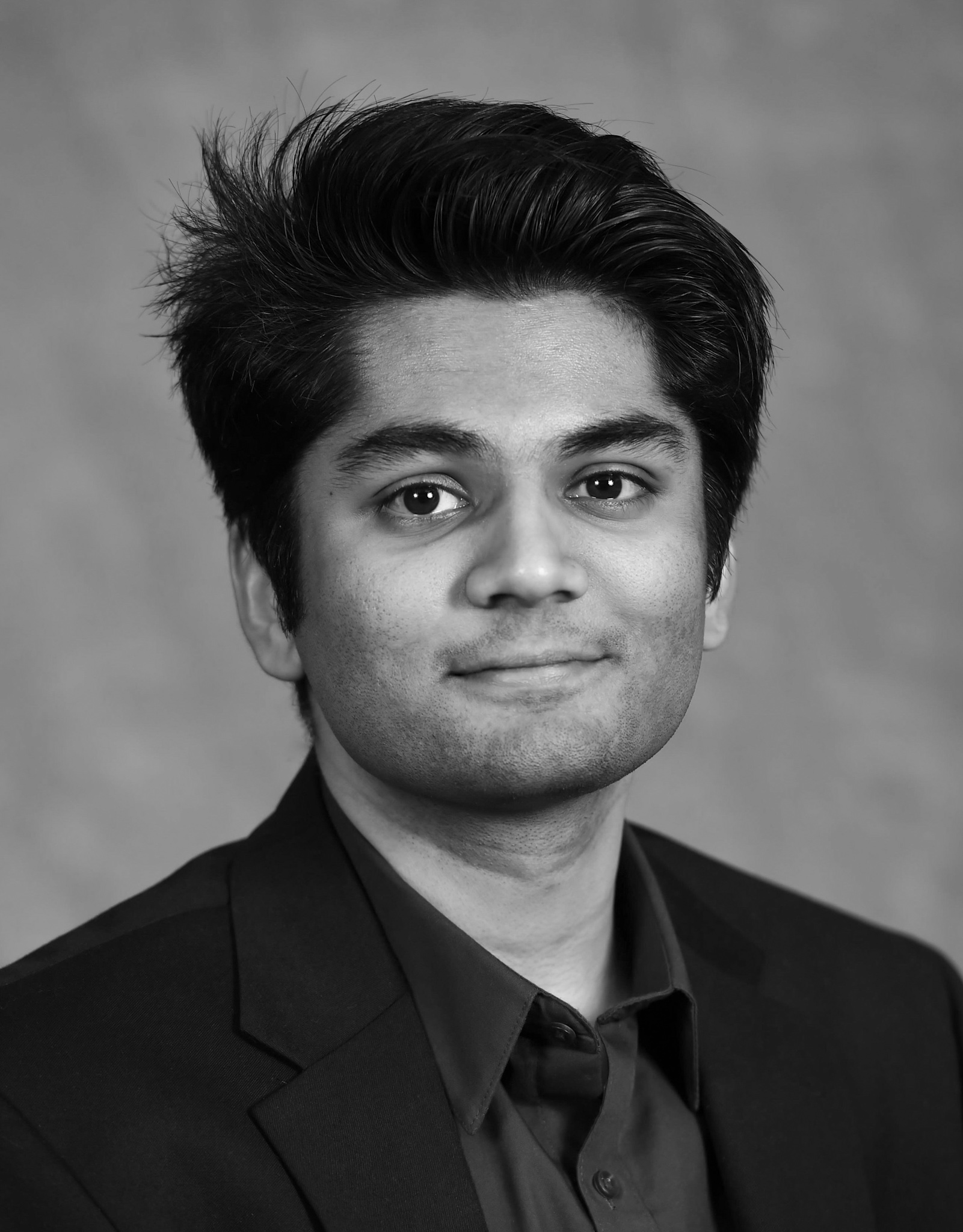}}

\begin{document}
\vspace*{4cm}
\title{(P)REHEATING EFFECTS OF A CONSTRAINED K{\"A}HLER MODULI INFLATION MODEL}

\author{ISLAM KHAN,~\textsuperscript{1,2} GUY WORTHEY,~\textsuperscript{1} AARON C. VINCENT~\textsuperscript{2,3}} 
\address{\textsuperscript{1}Department of Physics and Astronomy, Washington State University, Pullman, WA 99164, USA  \\
\textsuperscript{2}Arthur B. McDonald Canadian Astroparticle Physics Research Institute, Department of Physics, Engineering Physics and Astronomy, Queen's University, Kingston ON K7L 3N6, Canada \\
\textsuperscript{3}{Perimeter Institute for Theoretical Physics, Waterloo ON N2L 2Y5, Canada}}

\maketitle\abstracts{
In this talk, I discuss the effects, viability, and predictions of the string-theory-motivated K{\"a}hler Moduli Inflation I (KMII) potential, coupled to a light scalar field $\chi$, which can provide a possible source for today's dark energy density due to the potential's non-vanishing minimum. Although the model is consistent with the current measured Cosmic Microwave Background (CMB) data, tighter constraints from future observations are required to test the viability of the KMII potential with its minimum equivalent to the observed cosmological constant's energy density $\rho_{\Lambda_{\mathrm{obs}}}$. We implement a Markov Chain Monte Carlo (MCMC) sampling method to compute the allowed model parameter ranges and bounds on the inflaton's mass $m_{\phi}$ and reheating temperature $T_{\mathrm{reh}}$. Additionally, our lattice simulations predict stochastic gravitational-wave backgrounds generated during the inflaton oscillations that would be observable today in the $10^{9}$-$10^{11} \, \mathrm{Hz}$ frequency range. All the results and details will be included in our upcoming paper with the same title.}

\section{Introduction}
According to inflationary cosmology,~\cite{PhysRevD.23.347} the Universe underwent a period of rapid exponential expansion very early in its history. Among several others, inflation solves the horizon and flatness problems, and provides an attractive mechanism for explaining the observed structures in the Universe. An almost scale-invariant spectrum of primordial curvature perturbations imprinted in both the CMB and large scale structures also support an inflationary paradigm. In the simplest scenario, inflation is driven by a scalar field, namely the \textit{inflaton}, slowly rolling down its potential. At the end of inflation, it is generally assumed the inflaton coherently oscillates at the minimum of its potential, decaying and transferring its energy to a radiation-dominated plasma. This post-inflationary process during which the Universe gets repopulated with ordinary matter is known as \textit{reheating}. Rapid non-perturbative particle production effects can occur during the initial period of reheating, a period which is usually referred to as \textit{preheating}. The (p)reheating era is arguably the least understood epoch of the post-inflationary history.

String theory is a popular and promising candidate for the theory of quantum gravity and its applications to cosmology have regained interest over the last two decades, in particular, the development of string-theory-motivated inflation models. In this work, we consider a simplified version of the K{\"a}hler moduli inflation, referred to as the K{\"a}hler Moduli I Inflation (KMII)~\cite{Conlon:2005jm} (see also Blanco-Pillado \textit{et al.}~\cite{BlancoPillado:2009nw}), which has a non-vanishing potential minimum, providing a possible source for $\rho_{\Lambda_{\mathrm{obs}}}$. These models generally arise from the so-called Large Volume Compactification scenarios of Type IIB string theory. The potential's minimum is constrained by fixing its dimensionless free parameter $\alpha$ that characterizes the shape of the potential. In K\"ahler inflation models, $\alpha$ is related to the overall volume of the Calabi-Yau, the values of the other (stable) moduli, and couplings that are specific to a given compactification.~\cite{Conlon:2005jm} For simplicity, we consider a four-leg $\phi \phi \rightarrow \chi \chi$ quadratic interaction in all our analysis.

The KMII model provides one of the simplest descriptions of the physics within the context of modular inflation and it has not been ruled out by the current measured CMB data.~\cite{2020} It is also one of the simplest models with a non-vanishing minimum which can provide a possible source for $\rho_{\Lambda_{\mathrm{obs}}}$. The K\"ahler inflation model with a canonically normalized field is known as the ``K{\"a}hler Moduli II Inflation'' (KMIII) model,~\cite{Bond:2006nc} where the potential minimum takes large positive or negative values. In order to provide a source for $\rho_{\Lambda_{\mathrm{obs}}}$, the model must be consistent with observations when the potential minimum takes the value $V_{\mathrm{min}} \sim \rho_{\Lambda_{\mathrm{obs}}}$. The KMII model satisfies this condition, whereas the KMIII model does not.

\section{KMII Model}
\begin{figure}[!ht]
\centering
\includegraphics[width=12.7cm]{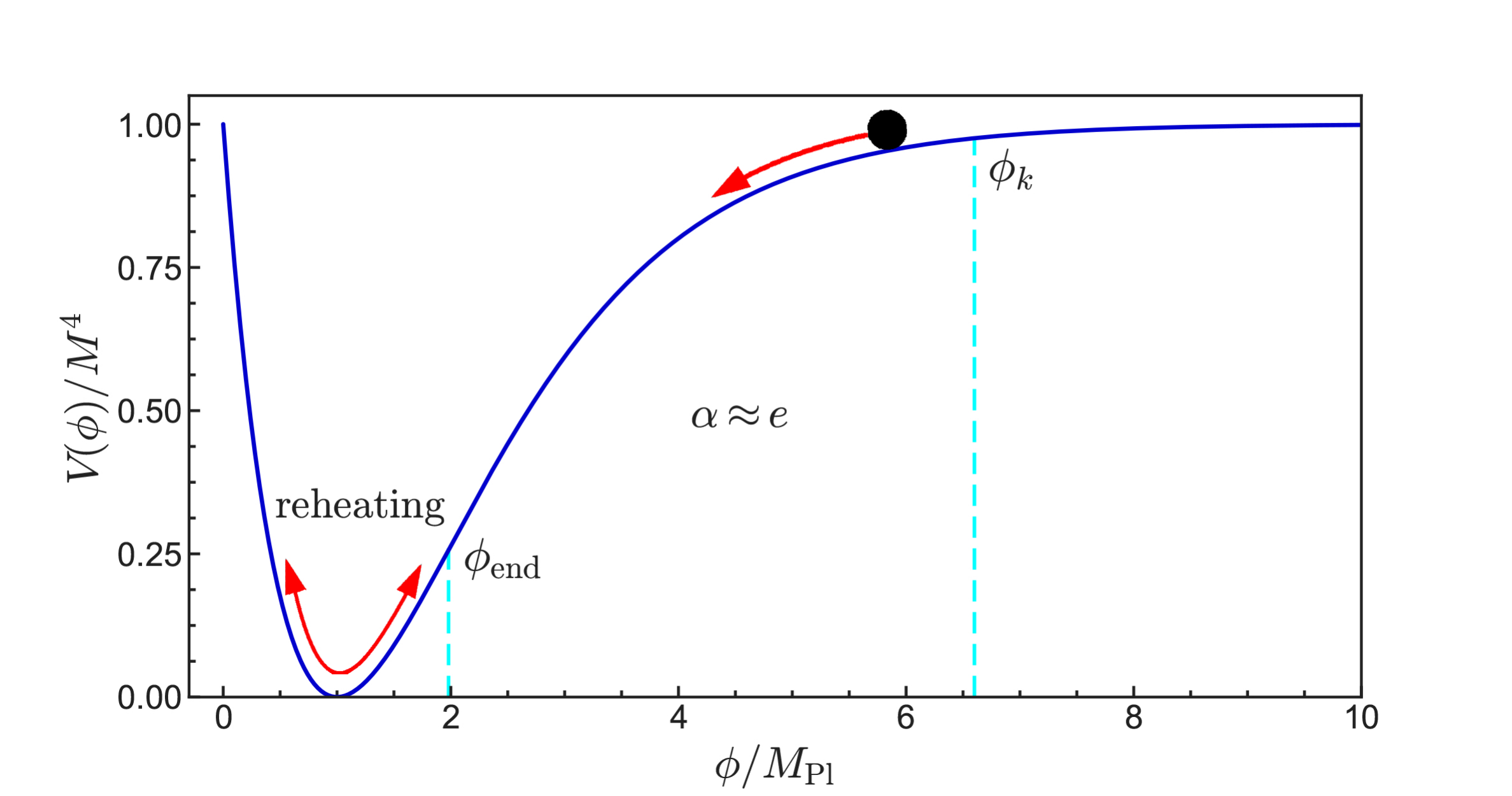}
\caption[]{The KMII potential for $1 - \alpha/e = 0$. The ball represents the inflaton slow-rolling down the potential. Reheating takes places at the minimum of the potential as the inflaton oscillates and transfers its energy to the other particles. The dotted vertical lines at $\phi_k \approx 6.6$ and $\phi_{\mathrm{end}} \approx 1.99$ correspond to the field values when the reference mode exits the horizon during inflation and inflation ends, respectively.}
\label{fig:kmiimodel}
\end{figure}

The KMII potential~\cite{Conlon_2006} is an example of a string theory-motivated inflationary potential. It was shown by Conlon \& Queved~\cite{Conlon_2006} that, when a large field limit is taken, the resulting inflationary potential can be simplified to $V =$ $M^4 \!  \left[ 1 - \alpha (\phi/M_{\mathrm{Pl}}) e^{-\phi/M_{\mathrm{Pl}}}\right]$, where $\phi$ is the inflaton field, $M$ is the energy scale, and $\alpha$ is a positive dimensionless parameter of the model. The KMII potential is shown in \fig~\ref{fig:kmiimodel} where $\alpha$ is fixed at $1 - \alpha/e = 0$. As shown by Martin, Ringeval, \& Vennin,~\cite{Martin:2013tda} $\alpha$ needs to be constrained at $\alpha \gtrsim 2.4095$ for inflation to successfully end by slow-roll violation. The potential has a minimum at $\phi = M_{\mathrm{Pl}}$ where it takes the form $V_{\mathrm{min}} = M^4(1 - \alpha/e)$. For our analysis, the inflaton is directly coupled to a light scalar field $\chi$ where it is assumed $\chi$ is short-lived and quickly decays to radiation. A four-leg interaction Lagrangian term $-g^2\chi^2\phi^2$, $g$ being the small coupling constant, is considered. The adopted model is then given by
\begin{equation}
V = M^4 \left( 1 - \alpha \frac{\phi}{M_{\mathrm{Pl}}} e^{-\phi/M_{\mathrm{Pl}}}\right) + g^2\chi^2\phi^2 \,. \label{eq:KMIIeq2}
\end{equation}
\section{Analysis and Results}
The three parameters of the adopted model as shown in \eq~\ref{eq:KMIIeq2} are $M$, $\alpha$, and $g^2$. A Markov Chain Monte Carlo (MCMC)~\cite{2013PASP..125..306F} sampling method, constrained by the latest release of \textit{Planck} CMB data,~\cite{2020} was implemented to compute the allowed ranges of the model parameters. A modified version of the HLattice code~\cite{2011PhRvD..83l3509H} was then used to study the (p)reheating dynamics of the adopted model.

\subsection{MCMC Analysis}
\begin{figure}[!ht]
    \centering
    \subfloat{{\includegraphics[width=8.6cm]{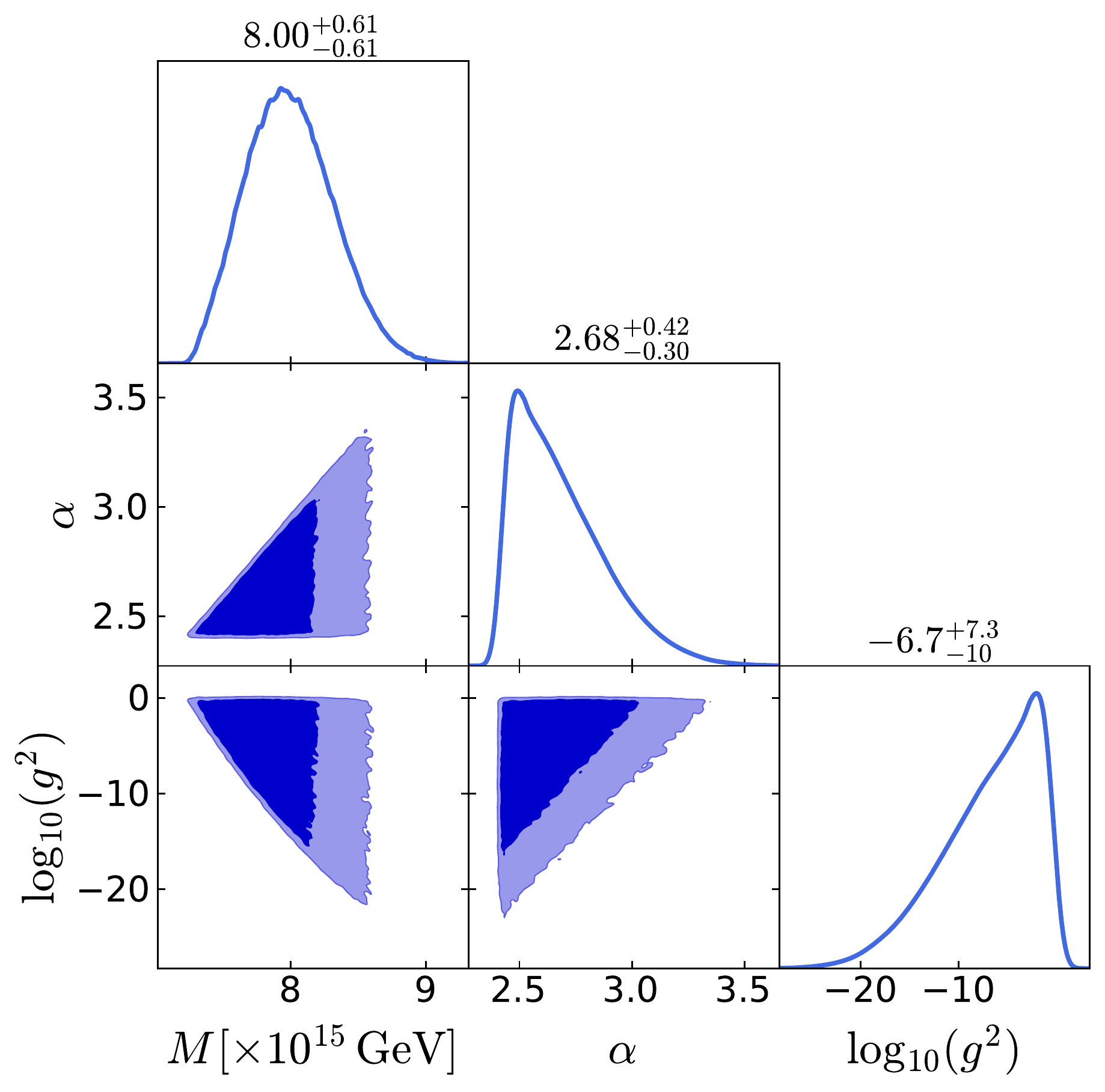}}}
    \caption{Triangle plot showing the one and two-dimensional posterior distributions of the adopted KMII model parameters $M$, $\alpha$, and $g^2$ from the MCMC sampling results. The marginalized probability distributions of the parameters are shown along the diagonal and the off-diagonal plots represent the two-dimensional distributions. The contours correspond to the $68\%$ and $95\%$ CL. The $68\%$ CL limits of the model parameters are shown on top of the diagonal plots.}
    \label{fig:mcmc}
\end{figure}

Within the slow-roll approximation formalism, we consider three slow-roll parameters $\epsilon$, $\eta$, and $\xi$ for quantifying inflation. These slow-roll parameters allow one to relate the model parameters to the CMB observables. They are defined by
\begin{equation}
\epsilon = \frac{M^2_{\mathrm{Pl}}}{2}\left(\frac{V'}{V}\right)^2 \, , \quad \eta = M^2_{\mathrm{Pl}}\frac{V''}{V} \, , \quad \xi = M^4_{\mathrm{Pl}}\frac{V' V'''}{V^2} \, , \label{slowroll1}
\end{equation}
\noindent where $V'$, $V''$, and $V'''$ are the first, second, and third derivatives of $V$ with respect to $\phi$. Inflation models can be constrained by the observed tensor-to-scalar power ratio ($r$), the scalar spectral index ($n_s$), and its \textit{running} ($n_{\mathrm{run}} = \mathrm{d \; \! ln} \; \! n_s/\mathrm{d \; \! ln} \; \! k$). At an arbitrary pivot scale $k = k_*$, they can be approximated as functions of the slow-roll parameters
\begin{equation}
r = 16\epsilon \, , \quad n_s = 1 - 6\epsilon  + 2\eta  \, , \quad n_{\mathrm{run}} = 16 \epsilon \eta - 24 \epsilon^2 - 2 \xi \, , \label{eq:slowroll1}
\end{equation}
and one can obtain the scalar power spectrum amplitude $A_s$ using
\begin{equation}
A_s = \frac{V(\phi_k)}{24 \pi^2 M^4_{\mathrm{Pl}} \epsilon} \, .  \label{eq:amplitude}
\end{equation}
\indent For the MCMC sampling analysis, the slow-roll parameter expressions (see \eqs~\ref{eq:slowroll1} and \ref{eq:amplitude}) were used corresponding to the adopted model. The constraints $\alpha > 2.4095$ and $g^2 < 1$ were applied during the analysis. Imposing flat priors for all three model parameters and initializing 100 walkers, the burn-in steps was set to 500 with a production run of 10000 steps. The marginalized posterior distributions of both the model and derived CMB observable quantities were computed. \fig~\ref{fig:mcmc} shows a triangle plot consisting of the one and two-dimensional posterior distributions of the adopted model parameters $M$, $\alpha$, and $g^2$ from the MCMC sampling results. The mass of the inflaton $m_{\phi}$ can be obtained from the curvature of the effective potential at its minimum given by $m^2_{\phi} = M^4 \alpha/ M^2_{\mathrm{Pl}}e$. The estimated allowed range of $m_{\phi}$ was computed to be $2.1 \times 10^{13} \, \mathrm{GeV} < m_{\phi} < 3.1 \times 10^{13} \, \mathrm{GeV}$ at $68\%$ CL. The reheating temperature $T_{\mathrm{reh}}$ is generally defined as the temperature of the Universe at the time of transitioning from the inflaton oscillation domination (reheating epoch) to radiation domination, under the assumption local thermal equilibrium has been reached~\cite{Albrecht:1982mp}. $T_{\mathrm{reh}}$ must be higher than the big bang nucleosynthesis energy scale ($T_{\mathrm{BBN}} \sim 1 \, \mathrm{MeV}$), and the upper bound of $T_{\mathrm{reh}}$ is constrained at $10^7$-$10^9 \, \mathrm{GeV}$ since higher temperatures can result in the production of unwanted relics such as gravitinos. $T_{\mathrm{reh}}$ corresponding to the adopted model can be estimated by
\begin{equation}
 T_{\mathrm{reh}} \sim \left( \frac{90}{g_* \pi^2} \right)^{1/4} \frac{g^2 M^2_{\mathrm{Pl}}}{\sqrt{8 \pi}M} \bigg( \frac{e}{\alpha}\bigg)^{1/4} \, . \label{eq:Treh2}
\end{equation}
\noindent Using \eq~\ref{eq:Treh2} and the MCMC sampling results, the lower bound on $T_{\mathrm{reh}}$ was estimated to be $T_{\mathrm{reh}} > 1.3 \times 10^{3} \, \mathrm{GeV}$ at $95\%$ CL.
\subsection{Lattice Simulation}
\begin{figure}[!ht]
\centering
\includegraphics[width=13cm]{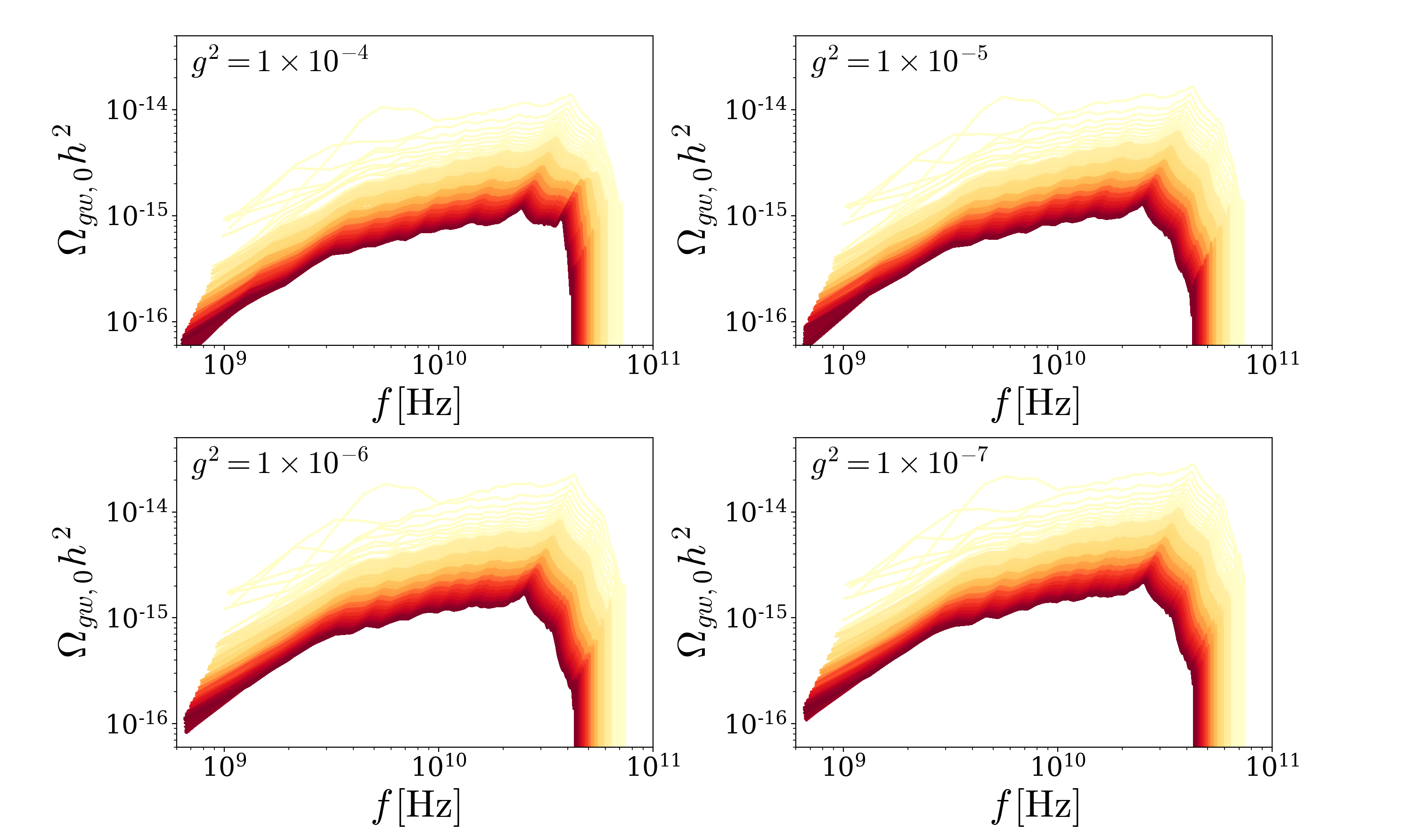}
\caption{Stochastic gravitational wave spectra generated due to inhomogeneities during the field oscillation in the adopted model for four different values of $g^2 = 10^{-4}, 10^{-5}, 10^{-6}, 10^{-7}$. The yellow represents $a = 1$ and red represents $a \approx 8$ (equivalent to about 2 $e$-folds). The results show no significant variation in the GW spectra as $g^2$ is varied. The energy conservation quantity $3H^2 M^2_{\mathrm{Pl}}/\rho_{\mathrm{tot}} - 1$ remained at least below $10^{-7}$ throughout in all the simulation runs.}
\label{fig:gw}
\end{figure}

A modified version of the HLattice code~\cite{2011PhRvD..83l3509H} was used to analyze the effects of (p)reheating in the adopted model. The HLattice parameters include the lattice box size at the start of the simulation ($L$) and box resolution ($n$). Energy conservation is enforced by requiring that the quantity $3H^2 M^2_{\mathrm{Pl}}/\rho_{\mathrm{tot}} - 1$ be sufficiently close to zero at all times, where $\rho_{\mathrm{tot}}$ is the total energy density of the system. The inflaton field was initially set to be homogeneous and the lattice simulation initial values of the inflaton field ($\phi_0$) and its kinetic energy ($\dot\phi_0$) were computed using the $\epsilon \geq 1$ condition. The lattice simulations were mainly used to compute the gravitational wave (GW) energy spectrum. The fractional energy of GW per $e$-fold is given by
\begin{equation}
\Omega_{\mathrm{gw}} = \frac{1}{\rho_{\mathrm{crit}}} \frac{d\rho_{\mathrm{gw}}}{d \, \mathrm{ln}f} \, , \label{eq:gw}
\end{equation}
\noindent where $f$ is the GW frequency, $\rho_{\mathrm{gw}}$ is the GW energy density, and $\rho_{\mathrm{crit}}$ is the critical density defined as $\rho_{\mathrm{crit}} = 3H^2 M^2_{\mathrm{Pl}}$ required for a spatially flat Universe. The GW energy spectrum in terms of present-day observables is denoted by $\Omega_{\mathrm{gw,0}}$ and it is obtained by replacing all the quantities in \eq~\ref{eq:gw} by today's observables (see Garcia-Bellido, Figueroa, \& Sastre~\cite{PhysRevD.77.043517} for details). With the $M$ and $\alpha$ parameters fixed at $M = 8 \times 10^{15} \, \mathrm{GeV}$ and $1 - \alpha/e = 0$, respectively, $\Omega_{\mathrm{gw,0}}$ was computed for $g^2 = 10^{-4}$, $10^{-5}$, $10^{-6}$, $10^{-7}$. The program parameters were set to $L = 0.3 H^{-1}_{\mathrm{ini}}$ and $n = 128$ in all the simulation runs. \fig~\ref{fig:gw} displays the results which shows no GWs are generated due to preheating instabilities within the parameter space; however, stochastic gravitational wave backgrounds (SGWBs) are generated due to inhomogeneities in the fields which would be observable today in the $10^{9}$-$10^{11} \, \mathrm{Hz}$ frequency range. Unfortunately, these frequencies are too high for any present or near-future GW observatories to probe. Varying $g^2$ does not significantly affect the characteristics of the GW spectra and the lattice simulation does not perform well when $g^2$ takes values $g^2 > 10^{-4}$. Varying the lattice spacing ($L$/$n$) within HLattice can significantly affect the SGWB amplitudes. However, the location of the peak frequency is largely independent of the non-physical simulation parameters.

\section*{Conclusion}
In an attempt to unify the two phases of accelerated expansions of our Universe within the context of modular inflation, this work studies the effects and viability of a simple inflation model known as the KMII model coupled to a light spectator scalar field $\chi$. Under the assumption that the total vacuum energy density of the Universe is zero due to some unknown symmetry, the dark energy density $\rho_{\mathrm{DE}}$, which can be attributed to the observed cosmological constant $\Lambda_{\mathrm{obs}}$, is modeled to come from the KMII potential's non-vanishing minimum. Unfortunately, the KMII model parameter $\alpha$ needs to be almost identical, but not equal to $e$, to achieve this result. This introduces a fine-tuning problem. In other words, considering $M = 8 \times 10^{15} \, \mathrm{GeV}$ and $\rho_{\mathrm{DE}} = 10^{-47} \, \mathrm{GeV^4}$, the $1-\alpha/e$ term in the KMII potential must be fine-tuned to about $110$ decimal places to achieve the desired result.

Our MCMC sampling analysis results estimate $2.1 \times 10^{13} \, \mathrm{GeV} < m_{\phi} < 3.1 \times 10^{13} \, \mathrm{GeV}$ at $68\%$ CL and $T_{\mathrm{reh}} > 1.3 \times 10^{3} \, \mathrm{GeV}$ at $95\%$ CL. The results also indicate $-0.14 < 1 - \alpha/e < 0.12$ at $68\%$ CL (see \fig~\ref{fig:mcmc}). If these aforementioned analyses, along with more precise observational data from future experiments, point toward $1 - \alpha/e = 0$ (or $V_{\mathrm{min}} = 0$), the energy density due to $\Lambda_{\mathrm{obs}}$ sourced from the non-vanishing minimum of the KMII potential will remain a possibility. On the other hand, if observations support $V_{\mathrm{min}} \neq 0$ instead, it would be ruled out. These analyses can also be extended to other types of inflaton interactions, \eg~trilinear interactions, and other inflation models that are consistent with observations and satisfy the $V_{\mathrm{min}} \sim \rho_{\Lambda_{\mathrm{obs}}}$ condition.
\section*{References}
\bibliography{main.bib}

\end{document}